# On the structure of protein-protein interaction networks

**Alun Thomas[1], Rob Cannings[2], Nicholas A. M. Monk[3] and Chris Cannings[3]**

1. Genetic Epidemiology, 391 Chipeta Way Suite D, Salt Lake City, UT 84108, USA
2. 10 Peterborough Drive, Sheffield, S10 4JB, UK
3. Centre for Bioinformatics and Computational Biology, and Division of Genomic Medicine, University of Sheffield, Royal Hallamshire Hospital, Sheffield, S10 2JF, UK

## Abstract

We present a simple model for the underlying structure of protein-protein pairwise interaction graphs that is based on the way in which proteins attach to each other in experiments such as yeast two-hybrid assays. We show that data on the interactions of human proteins lend support to this model. The frequency of the number of connections per protein under this model does not follow a power law, in contrast to the reported behaviour of data from large scale yeast two-hybrid screens of yeast protein-protein interactions. Sampling sub-graphs from the underlying graphs generated with our model, in a way analogous to the sampling performed in large scale yeast two-hybrid searches, gives degree distributions that differ subtly from the power law and that fit the observed data better than the power law itself. Our results show that the observation of approximate power law behaviour in a sampled sub-graph does not imply that the underlying graph follows a power law.

## Introduction

A collection of pairwise interactions within a set of proteins can be represented naturally as a graph in which vertices represent proteins and pairwise interactions are shown as edges. The collection of all interactions between the proteins of an organism is usually called the *interactome*. In this work we are concerned only with pairwise interactions, but we will use the term interactome to refer to this restricted set of interactions. The yeast two-hybrid (Y2H) system is a molecular technique for determining whether two proteins, or parts of proteins interact [1]. Y2H can be used to assay interactions between specific proteins or domains within proteins, or can be used to screen for interactions within pools or libraries of sequences [2]. Uetz *et al.* [3] and Ito *et al.* [4] have used Y2H to conduct systematic large scale searches of the interactome of the yeast *Saccharomyces cerevisiae*. The data from these screens can be represented graphically as an interaction network. Figure 1 shows the interaction network derived from the Uetz *et al.* screen [3].

The Y2H screens give a low density sample of the entire yeast interactome, as illustrated by the fact that there is little overlap between the two data sets. An alternative source of information is provided by the Pronet database (www.myriad-pronet.com). This contains data on the human interactome collated from the public domain, mainly from directed searches for interactions involving specific proteins. In contrast to the data on the yeast interactome, the coverage in Pronet is more patchy with almost full information where concentrated studies on particular pathways have been conducted, but sparse elsewhere. Similar data have been collated at the Database of Interacting Proteins (DIP; ref. 5).

**Figure 1:** The graph of the Uetz Y2H data [3] showing a single large component and many small ones. Protein labels and several self interactions have been omitted for clarity.

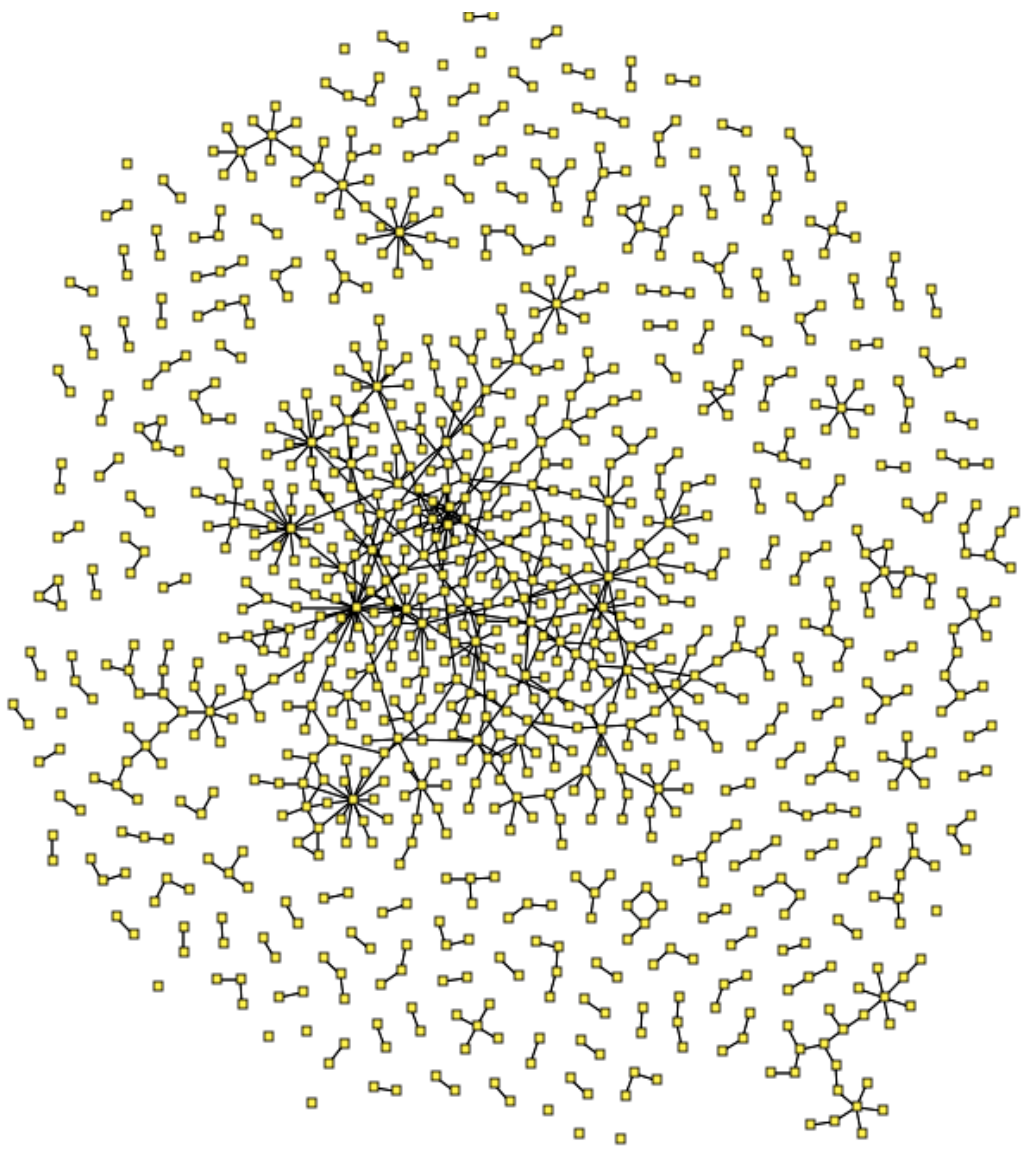

A number of analyses of interactome data have highlighted the apparent *scale free* behaviour of the graphs of observed protein-protein interactions [6,7]. The defining feature of scale free graphs is that the degrees $x$ of vertices are distributed according to a power law

$$f(x) \propto x^{-\gamma} \qquad x = 0,1,\ldots, \tag{1}$$

where $\gamma > 0$, so that a plot of log(degree) by log(frequency) shows a decreasing linear trend. For most examples of naturally-occurring power law distributions $\gamma$ is between 2 and 3. Power law behaviour has been reported for a number of diverse natural and engineered systems, such as metabolic networks and the World Wide Web [8–10]. The degree distributions of the data from the Y2H screens [3,4] are shown in Figure 5 and exhibit approximate power law behaviour. Given that the current data constitute a low-density sample of the entire yeast proteome, it is important to assess if the observed degree distributions accurately reflect the degree distribution of the full proteome.

The essence of the model we will present is the observation that parts of proteins, *domains*, contain sites into which complementary parts of other proteins can bind giving rise to the sort of interaction that is ascertained in a Y2H experiment [11]. We will refer to these complementary parts as the *positive* and *negative* aspects of a domain. Typically, a protein will contain several such sites, and the same aspect of a domain will be present in several proteins. We develop this model and explore its consequences on the structure of the interactome graph. The observed local structure of the human interactome provides strong evidence for our model. By invoking a simple stochastic model for the distribution of domains we derive the distribution of the degrees of vertices. This distribution is not a power law. However, we show that graphs derived by sampling from this model have a degree distribution that is closer to the power law and fits the observed Y2H data more closely than does the power law itself.

## Complete bipartite sub-graphs in protein-protein interaction networks

We suppose that each protein contains a set of *sites* and that each site is either the positive or negative aspect of some domain. Any pair of proteins one of which contains the positive aspect of some domain and one of which contains the complementary aspect of the same domain will be assumed to interact. This would appear as an edges in the interaction graph. In the context of the Y2H experiment, each aspect of the domain may appear as either a bait or a prey and the labelling of these as positive or negative is entirely arbitrary. The edges of the underlying graph are therefore undirected, although experimentally ascertained interactions can be given direction according to which protein was the bait and which the prey. The model allows a protein to interact with itself, or for a protein to have no sites and hence no interactions.

An immediate consequence of this model is that the graph of all protein-protein interactions is made up of complete bipartite sub-graphs–graphs comprising two disjoint sets of nodes in which each node in one set is connected to every node in the other set. Consider, for example, a particular domain for which the positive form is present in 3 proteins, *A*, *B* and *C*, and whose negative form is in 4 proteins *W*, *X*, *Y* and *Z*. The resulting interactome will then contain the complete bipartite graph on 3 and 4 vertices, usually written as $K_{3,4}$, in which each of *A*, *B* and *C* is joined to each of *W*, *X*, *Y* and *Z* as shown in Figure 2. In the human interactome, where there has been intense focus and near saturation sampling around several interesting proteins, we find considerable evidence for these complete and near complete bipartite structures, as illustrated in Figure 3.

**Figure 2:** The complete bipartite graph $K_{3,4}$.

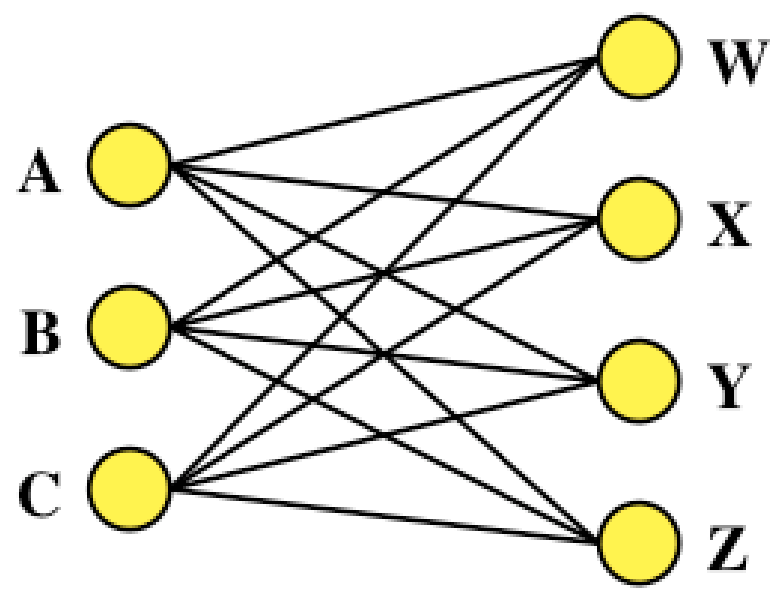

**Figure 3:** Examples of intersecting complete and incomplete bipartite graphs in the human interactome. Data was drawn from the Pronet database (www.myriad-pronet.com).

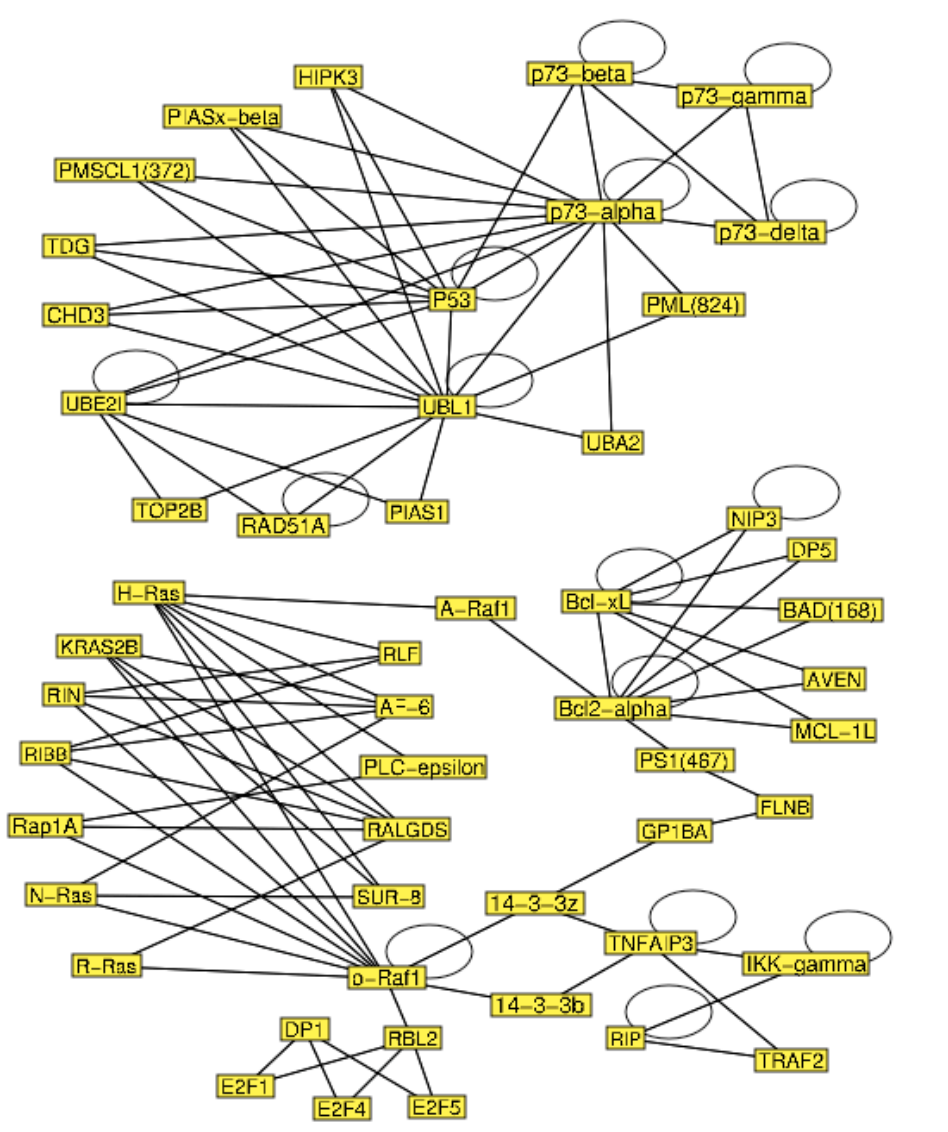

## Non-power-law degree distribution in a model network with random domain assignment

We shall assume that there are $n$ proteins and $m$ domains, each with a positive and negative form. A site may, therefore, be any of the $2m$ types $1^+$, $1^-$, $2^+$, $2^-$, ..., $m^+$, $m^-$. We assume the simplest possible random model, in which each of the $n$ proteins contains each of the $2m$ possible domains with constant probability $p$ independently of any other domains and independently of the contents of any other proteins. This is an example of the randomly coloured graph of Cannings and Penman [12]. Thus the number $X_i$, of sites that the $i$th protein has is distributed Binomially

$$X_i \sim \mathrm{Bin}(2m, p), \qquad (2)$$

and all the $X_i$ are independent and identically distributed. The average number of sites per protein is $\lambda = 2mp$. Our model is completely defined by the three parameters $n$, $m$ and $\lambda$. In the case of the yeast interactome, for instance, current data suggest that $n$, $m$ and $\lambda$ are in the region of 6000, 1000 and 1 to 2 respectively.

Let $Y_i$ be the number of interactions of the $i$th protein. If the $i$th protein contains $X_i = x$ sites, then any other protein $j$ will not connect to $i$ only if it does not contain any of the $x$ complementary domain aspects. So,

$$P(j \text{ does not connect to } i \mid X_i = x) = (1-p)^x. \qquad (3)$$

Since there are $n-1$ such other proteins, and domains are allocated independently, we have

$$Y_i | X_i = x \sim \mathrm{Bin}(n-1, 1-q^x), \qquad (4)$$

where $q = (1-p)$. Hence, the unconditional distribution of $Y_i$ is a Binomial mixture of Binomials

$$f(y) = \sum_{x=0}^{2m} {}^{n-1}C_y (1-q^x)^y q^{x(n-1-y)}\, {}^{2m}C_x p^x q^{2m-x}, \qquad (5)$$

for which we can also derive the inclusion/exclusion type expression

$$f(y) = {}^{n-1}C_y \sum_{i=0}^{y} {}^{y}C_i (-1)^i (q + pq^{(n-1-(y-i))})^m. \qquad (6)$$

Figure 4 shows the log(degree) by log(probability) plot of this distribution for $n = 6000$, $m = 1000$ and $\lambda = 1,2$, showing clear non-linearity. The curve for $\lambda = 2$ also shows multi modality as might be expected in a mixture. While we focus here on the form of the degree distribution, we note that other features of the interactome can be derived under this model. For example, the expected number of triangles as $n \to \infty$ with $n = Km$ for fixed $K$ and $\lambda$, is $K^3\lambda^4(3 + 3\lambda + \lambda^2)/48$ (which, for $K = 6$ takes the value 31.5 for □ = 1 and 156 for □ = 2). Consequently, the clustering coefficient of the interaction graph is asymptotically zero.

**Figure 4:** Log-log plot of the distribution of vertex degrees in the modelled interactome with 6000 proteins, 1000 domains and an average of 1 or 2 domains per protein, shown as solid and dotted lines respectively.

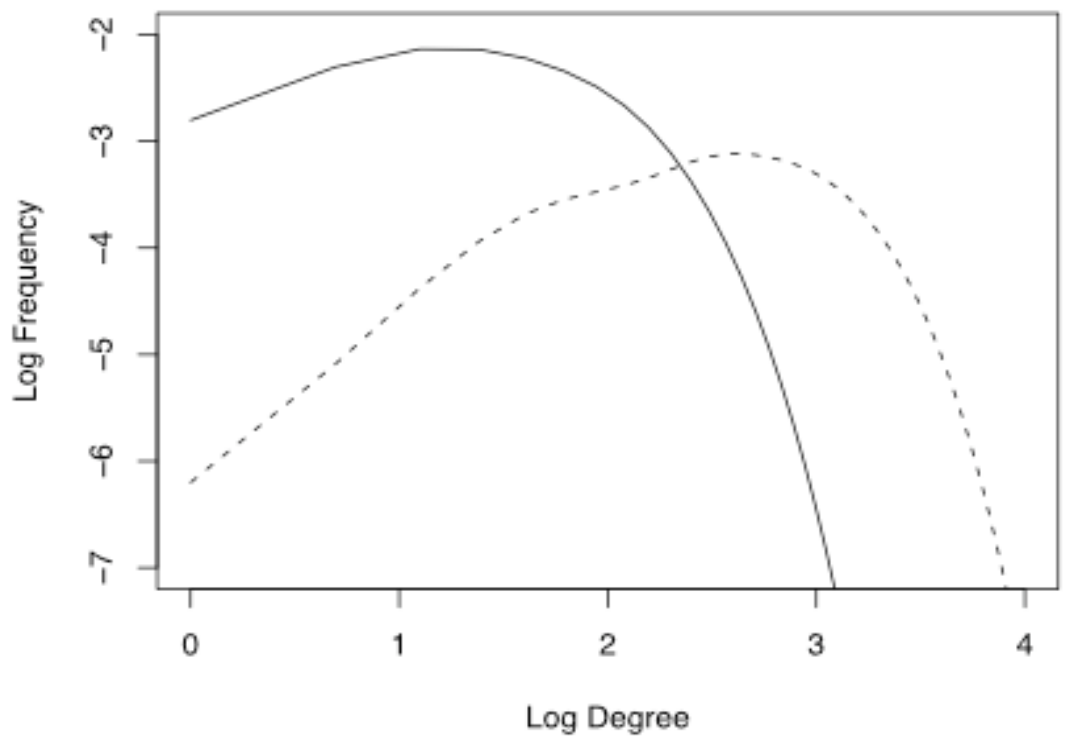

## Degree distribution of sampled sub-graphs

The above model predicts the structure of a complete interactome. However, the data provided by the Y2H screens [3,4] are samples from the complete yeast interactome. Although each screen could in principle have found all interactions, experimental conditions are such that not all have been found [13]. Typically there is competition for resources in the mating pools and on the growing media between yeast colonies. Also, the strategy used by both studies of looking for interactions between complete proteins is known to be less sensitive than protocols that use partial sequences. The accumulated data for yeast in the DIP currently has over 15000 interaction connecting over 4700 proteins.

To model the type of sampling used in the Y2H screens, we simulated complete interactomes from our model with (*n*,*m*,□) = (6000,1000,1) and sampled interactions from them according to the following scheme. A total of 450 proteins were sampled at random with uniform probabilities from the set of 6000. For each protein we sampled with replacement a Geometrically distributed number of neighbours in the interaction graph. The mean number of neighbours sampled was 5. Sampled proteins for which either no neighbours were available or no neighbours were sampled were discarded to reflect the nature of the Y2H procedure. The graphs obtained had approximately the same number of vertices and edges as the Uetz *et al*. data set [3]. The degree distributions of the sampled sub-graphs were calculated and averaged over 100 independent interactome simulations and 100 sub samplings for each independent interactome. The resulting degree distribution is shown in Figure 5. While the distribution is more like a power law than that for the underlying complete interaction graphs, it does exhibit clear non-linearity. However, the distribution is a better fit to both the Uetz *et al*. and Ito *et al*. data than is a straight line (power law).

We also fitted our model to all the current yeast data from the DIP. In this case the number of proteins and domains in the underlying graph were again 6000 and 1000. To reflect the accumulated sampling from several studies the number of proteins sampled was increased to 1500 and we assumed that all of the neighbours of each protein sampled were ascertained. The mean number of domains per protein was taken to be 2. The resulting degree distribution is shown in Figure 6. Again the fit of the model to the data is better than the power law.

**Figure 5:** Log-log plot of the observed and expected distribution of vertex degrees in the Ito [4] and Uetz [3] yeast Y2H experiments. The Ito and Uetz datasets are plotted in black and blue, respectively; a straight line (power law) fit is shown as a dotted line. The distribution obtained by sampling from our model with 6000 proteins, 1000 domains and an average of 1 domain per protein is plotted in red.

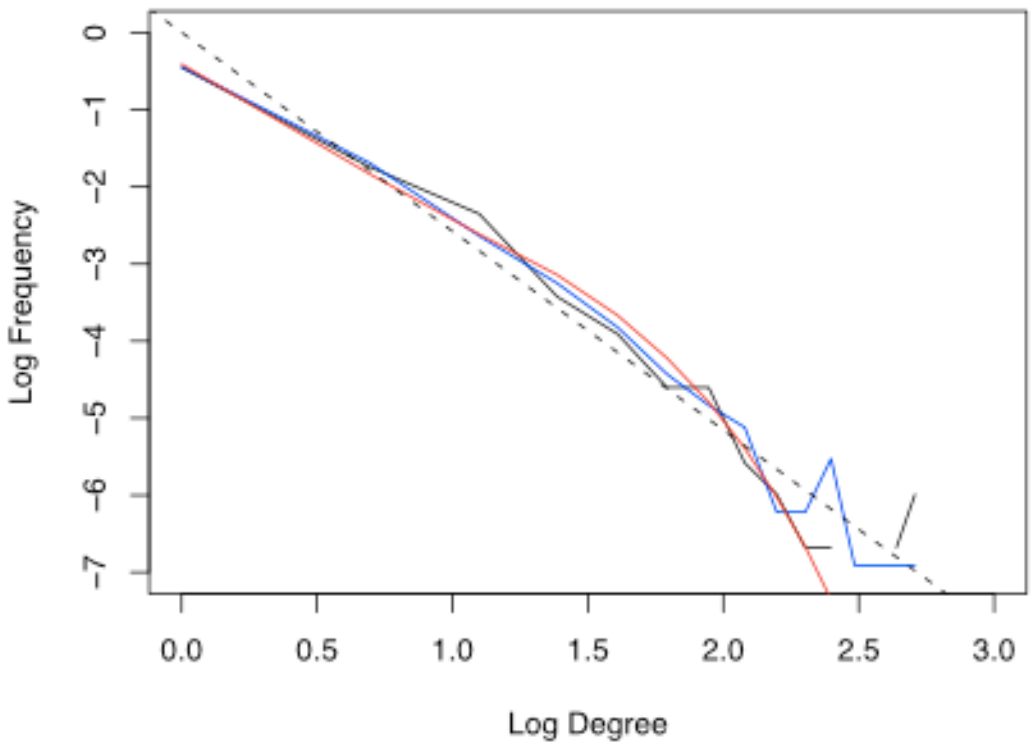

**Figure 6:** Log-log plot of the observed and expected distribution of vertex degrees in the current DIP yeast interaction database. The DIP dataset is plotted in black. The distribution obtained by sampling from our model with 6000 proteins, 1000 domains and an average of 2 domains per protein is plotted in red. A straight line (power law) fit is shown as a dotted line.

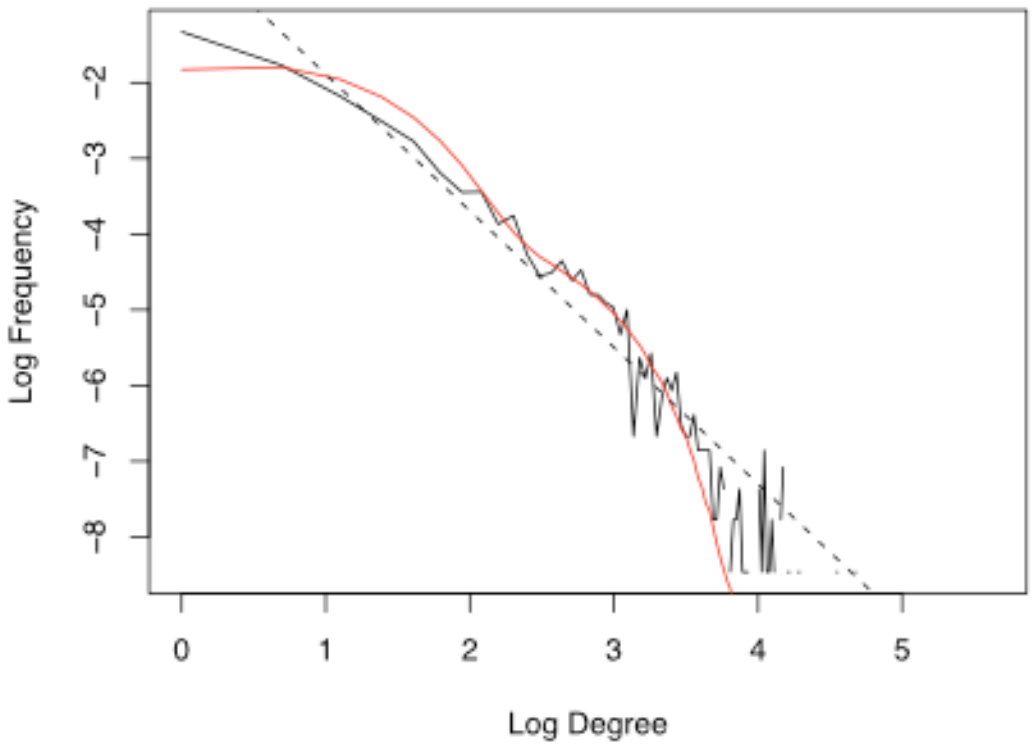

## Discussion

We have proposed a model for the structure of interactome graphs and for sub-graphs obtained by the type of sampling employed in Y2H screens. There is clear evidence supporting this model in densely sampled portions of the human interactome and we have shown that sampling from our model yields degree distributions very like those resulting from Y2H screens of the yeast interactome. Indeed, the degree distributions predicted by the model fit the data better than do power law distributions. At the current level of sampling, the deviation from linearity seen in our model and in the Y2H data sets is slight. Our findings thus show that observation of an approximate power law degree distribution resulting from sampling of a graph does not imply that the underlying graph exhibits a power law distribution. However, as the sampling density is increased, our model predicts increasingly significant deviation from a power law.

This conclusion is strengthened by the fact that performing the same sampling from a classical random (Erdös-Rényi; ref. 14) graph with the same number of vertices and approximately the same number of edges also gives an approximate power law for most of the range, but with a lighter tail (data not shown). Furthermore, Barabási and Albert have shown that a particular form of sampling (based on preferential attachment) of a fixed number of nodes produces a power law degree distribution at low density, but not at higher density [15]. We note also that the long right hand tail observed in the degree distributions is somewhat misleading. There are very few outliers, and for all but one of the proteins reported to have more than 15 interactions, these occurred with the protein always being the bait or always being the prey in the Y2H assay. This suggests that the length of the right tail may be due to experimental artifacts.

Several applications of our model suggest themselves immediately. One is to use the predicted bipartite structure of sub-graphs to infer the existence of interactions not yet detected experimentally. For instance, Figure 3 strongly suggests that o-Raf1, PLC-ε, RALGDS, AF-6, RLF and SUR-8 contain a motif that interacts with a complementary motif in R-Ras, Rap1A, KRAS2B, RIN, RIBB, N-Ras and H-Ras. This would imply that for instance RLF and AF-6 should interact with Rap1A and R-Ras in order to complete the bipartite graph. If Y2H experiments specifically directed at finding these interactions do not give positive results this would raise interesting questions about the structure of the proteins involved, or about whether Y2H experiments systematically miss interactions. Conversely, we should also be able to remove false edges, particularly for vertices of high degree. As sampling gets closer to saturation we should see greater departure from linearity in the degree distribution. Observation of such a departure would indicate a good time to switch from random sampling to directed experiments.

By fitting the model to currently available data we can address questions about the number and nature of domains in a quantitative manner. The fits of the model shown in Figures 5 and 6 were performed by trial and error and using parameters estimated from other sources, and serve only as a proof of concept. We are currently developing a systematic maximum likelihood approach for the problem. This is a Markov chain Monte Carlo method which simulates likely underlying graphs from our model given the observed data in much the same way as image restoration techniques simulate true pictures given erroneous data [16]. In a similar fashion we should be able to clean up our picture of the interactome.

Having a clearly specified model gives us a baseline from which departures are more clearly seen. For instance in Figure 3 the configuration of the proteins p73-α, p73-β, p73-γ, and p73-δ is a complete sub-graph, in which all vertices are joined to each other, rather than a complete bipartite sub-graph. Moreover, each of the proteins interacts with itself. This suggests that each of the proteins contains a single symmetrical motif that interacts with itself. This is indeed the case; the proteins are splice variants, each containing the sterile α-motif that is known to mediate homodimerisation [17, 18]. Our model can be extended straightforwardly to incorporate such symmetric domains.

## Acknowledgments

N.A.M.M. gratefully acknowledges the support of the University of Sheffield (J.G. Graves Research Fellowship).